\documentclass{aastex}
\usepackage{graphicx}

\newcommand{\etal}{{\it et al.\ }}

\begin{document}

\title{The CFEPS Kuiper Belt Survey: Strategy and Pre-survey Results}
\author{
R. L. Allen
\footnote{
Based on observations obtained at the Canada-France-Hawaii Telescope (CFHT) which is operated by the National Research Council of Canada, the Institut National des Sciences de l'Univers of the Centre National de la Recherche Scientifique of France, and the University of Hawaii, and on further observations obtained at the ESO Telescopes at La Silla and Paranal Observatories. In addition, based on data taken using ALFOSC, which is owned by the Instituto de Astrofisica de Andalucia (IAA) and operated at the Nordic Optical Telescope under agreement between IAA and the NBIfAFG of the Astronomical Observatory of Copenhagen, observations collected at the Centro Astron—mico Hispano Alem‡n (CAHA) at Calar Alto, operated jointly by the Max-Planck Institut fŸr Astronomie and the Instituto de Astrof'sica de Andaluc'a (CSIC), and observations from the Hale Telescope, Palomar Observatory, operated as part of a collaborative agreement between the California Institute of Technology, its divisions Caltech Optical Observatories and the Jet Propulsion Laboratory (operated for NASA), and Cornell University. Also, from observations taken as a visiting astronomer, Kitt Peak National Observatory, National Optical Astronomy Observatory, which is operated by the Association of Universities for Research in Astronomy, Inc. (AURA) under cooperative agreement with the National Science Foundation.
},
%
B. Gladman \\
University of British Columbia, CANADA \\
J-M. Petit, P. Rousselot, O. Mousis \\
Observatoire de Besan\c{c}on, FRANCE\\
J.J. Kavelaars \\
HIA/NRC, CANADA \\
A. Campo Bagatin, G. Bernabeu, P. Benavidez \\
Universite de Alicante, SPAIN \\
J. Parker\\
Southwest Research Institute, USA\\
P. Nicholson\\
Cornell University, USA\\
M. Holman \\
Harvard-Smithsonian Center for Astrophysics, USA\\
A. Doressoundiram,  \\
Observatoire de Paris-Meudon, FRANCE\\
C. Veillet\\
CFHT, USA\\
H. Scholl, G. Mars\\
Observatoire de la C\^{o}te d' Azur, FRANCE\\
}

\begin{abstract}
We present the data acquisition strategy and characterization procedures
for the Canada-France Ecliptic Plane Survey 
(CFEPS), a sub-component of the Canada-France-Hawaii Telescope Legacy Survey.
The survey began in early 2003 and 
as of summer 2005 has covered 430 square degrees of sky
within a few degrees of the ecliptic. Moving objects beyond the orbit of Uranus are
detected to a magnitude limit of $m_R$=23 --- 24 (depending on the image quality).
To track as large a sample as possible and avoid introducing
followup bias, we have developed a 
multi-epoch observing strategy that is spread over several years.
We present the evolution of the uncertainties in ephemeris position and orbital elements
as the objects progress through the epochs.
We then present a small 10-object sample that was tracked 
in this manner as part of a preliminary survey starting a year before the main 
CFEPS project.

We describe the CFEPS survey simulator, to be released in 2006,
which allows theoretical models of the Kuiper Belt to be compared
with the survey discoveries since CFEPS has a well-documented 
pointing history with characterized detection efficiencies as 
a function of magnitude and rate of motion on the sky.
Using the pre-survey objects we illustrate the usage of the simulator
in modeling the classical Kuiper Belt.
\end{abstract}

\section{Introduction}

Exploring the Kuiper belt allows us to study the process of planet 
formation and gain insight into the evolution of the outer Solar 
System.  
By studying the dynamical environment of these distant planetesimals, we can provide limits to planetesimal accretion and erosion models \citep{Stern97, Stern97b, Kenyon99}. 
Clues to the outer Solar System's evolution 
are hidden in the relative populations of the various components
of the trans-neptunian population \citep{Ma95, Ma96, Chiang02, Levison03, Gomez05}. 
A census of these dynamical populations allows us to identify potential source regions for the origin of short-period comets 
\citep{Levison97}. However, an accurate picture of the Kuiper belt is difficult to determine in the
presence of extreme observational biases related to flux and orbit 
determination.

The biggest repository 
of Kuiper belt orbits is the Minor Planet 
Center\footnote{The Minor Planet Center database is online at 
http://cfa-www.harvard.edu/iau/mpc.html}, 
whose lists include all reported TNOs with linked
observational arc-lengths of two nights or more. 
While there are over 1000 Trans-Neptunian objects (TNOs) in the Minor Planet 
Center (MPC) database, over 300 of these objects are currently 
considered ``lost'' as their ephemeris uncertainty is now 
over 1 degree. 
The reasons why some objects are lost and other objects are not 
vary --- followup observations have generally been done on an 
ad-hoc basis, so followup success depends on telescope scheduling,
the brightness of the TNO, and how closely the initial orbit 
estimate matches the actual orbit. 
The orbit assumptions can be an important ``invisible'' bias, as these 
generally tend to place newly discovered TNOs on orbits similar to 
previously known TNOs. 
In addition, the followup observations are never systematically
documented, so one does not know if a given object was lost because
it was never followed (in which case the orbit listed in the database
may be correct) or attempted and not found (in which case the database's
orbit is certainly {\it incorrect}!).
The combination of these factors leads to possible biases in the 
orbit distributions in the MPC database. 
Out of the $\sim$1000, only about 300 MPC TNOs have orbits well-determined 
enough for use in dynamical models, but even these lack information on 
the discovery/recovery conditions. 
Published individual surveys with characterized detection conditions, 
including some followup details, contain 
only tens of objects at most, providing 
a very small sample size. 

TNOs are loosely classified according to their orbits, into resonant, 
classical, scattered disk and extended scattered disk populations 
(see \citet{LJ02, Gladman02, Gladman02b}. 
The extended scattered disk population was only relatively recently 
discovered
\citep{Gladman02}. 
Notably, if 2000 CR105 had not been tracked 
to a longer arc within its initial opposition
outside its discovery survey, it 
would have appeared as a normal scattered disk object. 
Thus, the relative population of extended scattered disk objects cannot be 
determined without a knowledge of the tracking rate of discovered TNOs 
and a tracking program free from orbital biases. 
It is also important to have a large sample size, in order to be able 
to measure the population fraction of rare groups such as the extended 
scattered disk,  Neptune Trojans \citep{Chiang05},
or TNOs in distant mean-motion resonances; 
these classes appear to form only tiny fractions of a flux-limited
trans-neptunian catalogue.

The population densities of TNOs in each mean-motion resonance with Neptune, 
and the distribution of objects within these, help constrain models 
of Solar System evolution.
For example, the relative fraction of plutinos (in the 3:2 resonance) versus 
``twotinos'' (2:1 resonance) might be used to determine the speed and 
extent of an outward migration of Neptune \citep{Ma95, Ida00, Chiang02}. 
In the 2:1 resonance, the speed of  Neptune's migration and the initial 
conditions of the planetesimal disk affect the likelihood of capture 
into either leading or trailing libration islands \citep{Murray05}. 
The longitudinal asymmetries caused by a resonant orbit dramatically alter the 
chances of finding resonant TNOs when pointing at particular ecliptic 
longitudes; the pointing history and search conditions of the survey are crucial 
for de-biasing from observed population fractions to the 
intrinsic ratios, even when objects are {\it not} discovered. 
This survey work has been designed to 
avoid these issues, and provide an exceedingly well-understood survey of the
Kuiper Belt that comes with the tools needed to compare models 
with the survey's results.

\subsection{The CFEPS survey}

The Canada-France Ecliptic Plane Survey (CFEPS) is a part of the 
Canada-France-Hawaii Telescope Legacy Survey (see 
http://cfht.hawaii.edu/Science/CFHTLS/ for more details).
The Legacy survey is divided into 3 components, with each component
serving more than one science goal. 
As of July 31 2005,
the Very Wide component of the Legacy Survey (LS-VW) has covered 
approximately 430 square degrees near the ecliptic, in multiple filters 
and with a widely-spaced exposure timing.
The CFEPS Kuiper Belt study is the primary science driver for the
LS-VW, and drives the exposure timing for this component of the survey. 
Each field is acquired in multiple filters to serve other science 
goals (including galactic structure, stellar populations, and 
high-redshift quasar searches). 
However, the filter sequencing and exposure times are driven by the desire
to detect TNOs and provide some portion of 
the tracking observations needed to determine their orbits for the CFEPS 
study. 
The goal of CFEPS is to obtain an unbiased census 
of Kuiper belt's various components. 
The survey will be well-characterized in terms of completeness and 
magnitude limits at each pointing, but also in the orbital tracking of 
each detected TNO. 
With the CFEPS sample, we intend to provide an orbital database 
suitable for dynamical modeling of the Kuiper Belt. 

In order to refine the CFEPS strategy,
we conducted a small presurvey prior to the full onset of 
the project. 
This paper describes the CFEPS presurvey and how it affected the overall project 
design. 
The presurvey covered 
a 6.5-square degree patch of sky and resulted in the discovery of 13 TNOs;
ten of these were within the expected flux limit of CFEPS, 
and were followed in an intensive observational 
campaign. 
In addition, we describe the CFEPS simulator 
for comparing model Kuiper belts against our survey observations.

\section{Observations}

While this team has had significant prior experience in detecting 
TNOs \citep{Gladman01, Petit05, Allen01}, 
the presurvey observations detailed here were conducted specifically 
as a test of our design for the CFEPS observation strategy and a trial 
of our methods for obtaining and reducing the data. 
The queue programming, observation strategy (particularly optimum 
observation timing), and our ability to subsequently track the
discovered objects was of particular importance. 

\subsection{The Observational Plan}

Given the CFEPS goal of producing a characterized TNO orbital database,
the discovery observations must be characterized 
for limiting magnitude and completeness levels, and the recovery 
efficiency should be unbiased by orbital assumptions. 
In order to help attain this goal, CFEPS has 
some recovery observations built into the survey plan. 
These recovery observations are not targeted to specific ephemerides 
and thus are not affected by orbit assumptions. 
TNOs may still fail to be recovered due to moving in front of a bright star, 
into a gap in the camera, shearing out of the field coverage or simply 
being below the limiting magnitude of the observations, but not 
directly because of orbit assumptions. 
The generalized observational sequence is as follows: 

\begin{itemize}
\item \underline{Discovery} consists of a triplet of images. Each 
pointing is imaged three times, with each image separated by 45 minutes 
to 1 hour. 
These images are searched for new TNOs and characterized for completeness 
and magnitude limits, using methods described in \citet{Petit04}.
\item \underline{Nailing} is a single image of the discovery pointing, 
taken within four nights of the discovery triplet (either before or after) .  The
close proximity of the discovery triplet allows this single image to be used
to identify (aka ``nail'') the objects found in the discovery images.
\item \underline{Checkup} consists of another triplet of images, taken 
within a few months (before or after) of the discovery images. 
This is primarily intended to improve the ephemeris of the TNO for 
one-year recovery. 
\item \underline{One-year Recovery} observations are similar to discovery - 
a triplet of images on one night, combined with a single image within the 
same dark run. 
These observations add to the total observed arc, generating a 
well-determined ephemeris, but the arc length is still insufficient to 
allow precise determination of the orbital parameters. 
\item \underline{Orbit Classification} 
observations are again similar to discovery or one-year recovery. 
With these observations at least two years after discovery, the orbit 
is finally well-determined (usually to within 0.01~AU in semi-major 
axis), permitting the dynamical classification of the TNO. 
\end{itemize}

This was considered the minimum tracking strategy for CFEPS. 
The presurvey provided the opportunity to 
determine the optimum timing for each of these steps in order to minimize 
object losses and maximize observational efficiency. 
Many of the TNOs discovered in the CFEPS presurvey received much more 
extensive tracking observations, relatively densely spaced in time. These TNOs serve as test cases of when observations are maximally useful and possible within the CFEPS survey format. 

\subsection{The Presurvey Observations}

\begin{figure}[htbp]
\includegraphics*[width=7.5in]{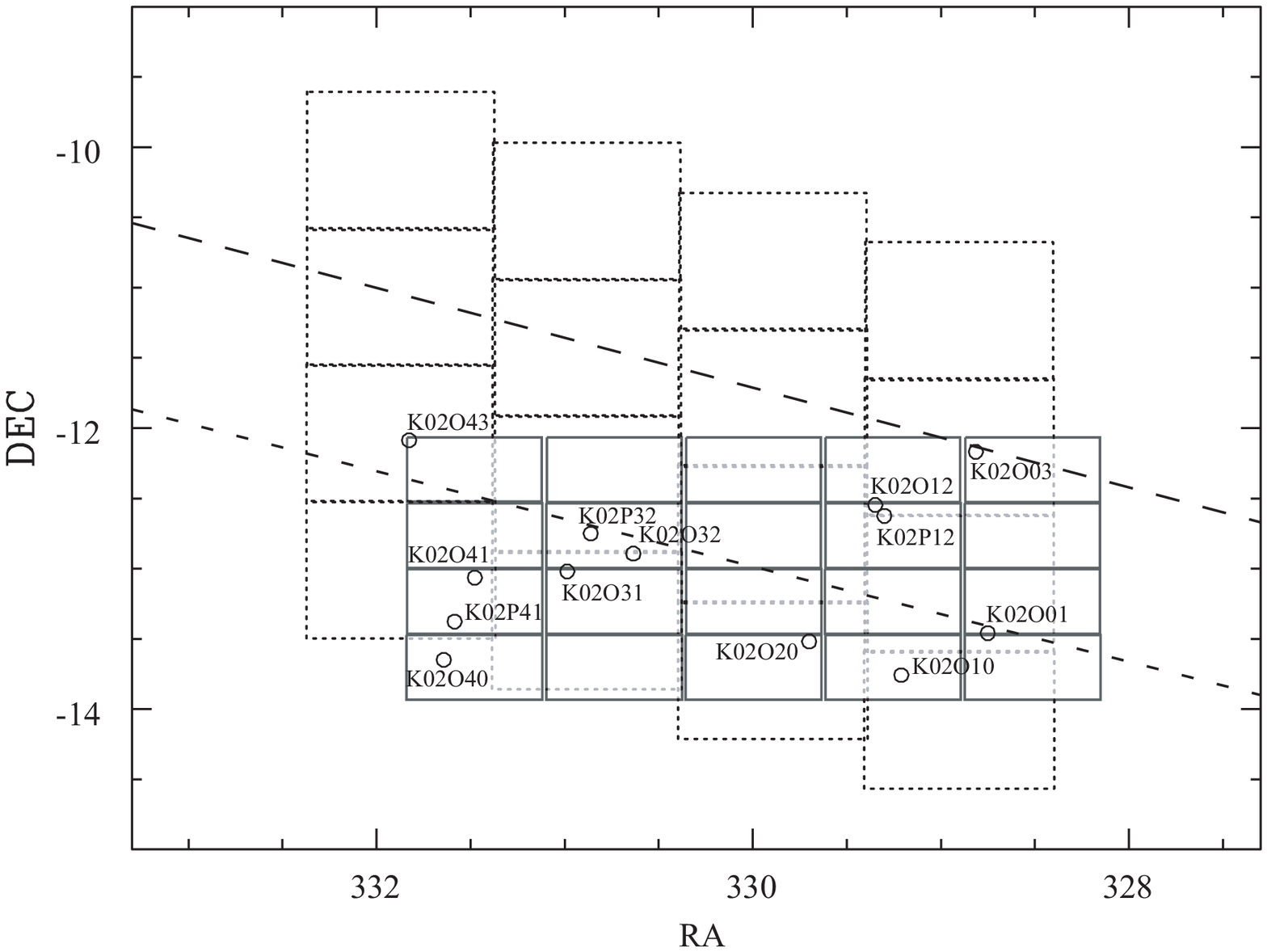}
\caption{Field layout for presurvey (solid outlines) and CFEPS (dotted outlines) discovery images. The upper dashed diagonal line through the plot is the location of the ecliptic plane, the lower dashed diagonal is the location of the invariable plane. The locations of objects detected within the presurvey are marked on each field. The true aspect ratio is such that the CFEPS Megacam fields are square.}
\label{fieldlayout}
\end{figure}

The presurvey TNO discovery images were obtained at the 
Canada-France-Hawaii Telescope on August 5, 2002 using the CFH12K camera 
and the Mould R filter.  
These discovery images consisted of a set of 20 different field pointings, 
each pointing imaged three times during the night. 
The CFH12K camera has a $28\arcmin$ x $42\arcmin$ field of view, 
which allowed us to cover a contiguous 6.5 square degree patch of sky. 
The layout of these field pointings is shown in Fig.~\ref{fieldlayout}. 
One two-minute image at each of the 20 pointings was taken, and then 
the entire set of 20 pointings was imaged a second time, and
then a third time. 
This produced a 1.5-hour time separation between each image in the triplet. 
Given typical rates of motion (all known TNOs move between 1.5 and 5 "/hr
at opposition) this spacing is more than adequate to detect the moving TNOs.
This was the natural result of cycling through the sequence of 
20 pointings, with each image having a 2 minute exposure time followed 
by a 2 minute readout time. 
A 2 minute exposure time for these discovery images ensured 
the overheads associated with the CFH12K camera readout did not result in a less than
50\% duty cycle.
The presurvey observations provided valuable insight into the efficiencies 
of the telescope pointing system, guide star acquisition, camera overheads, 
and overall observing efficiency for the types of sequenced observations 
that are part of the CFEPS project.

After searching the fields with our moving object detection pipeline
\citep{Petit04}, thirteen TNOs were discovered.
They ranged in brightness from $m_R=22.3-23.9$ 
(see Table~\ref{objectstable} for a list), 
and heliocentric distances estimated to be between 40 to 60~AU.
Because exposure times were two minutes and
the discovery conditions were good -- average seeing 
was 0.7\arcsec\ with very clear skies -- the faintest presurvey TNOs are
fainter than the average expected from the main CFEPS survey (which uses 
80-sec exposures in median seeing 0.9").

The exposure time and filter was the same for the nailing image as for 
the discovery triplet. 
The nailing image for these twenty fields was not obtained until 
August 8 for some fields, or August 9 for others, 
due to bad weather at CFHT. 
We found that this four-night delay to be slightly problematic, as the 
larger change in TNO position makes objects harder to detect in 
the nailing image as their on-sky error ellipse becomes 
{\em much} larger. 
Three of the TNOs were not visible on their nailing images due
to problems like these.

Some followup observations for all 13 TNOs (including those which did not get nailing measurements) were obtained at 
a variety of telescopes, utilizing significant resources to ensure 
as many objects as possible were tracked 
with as high a cadence as feasible, especially within the 
discovery opposition.
Table~\ref{objecttracking} contains details
of the observational campaign. 
All thirteen presurvey TNOs were re-observed 
one month after discovery,
at either the Nordic Optical Telescope, the VLT, and/or at CFHT. 
These checkup observations consisted again of a triplet of images. 
At CFHT, the checkup images were of the same pointings as the initial 
discovery images (as is the case for CFEPS), but the other telescopes 
were targeted to the predicted position of each object. 
Telescopes 4-m and up (such as the VLT and CFHT) were needed for some of 
these 1-month checkup observations.
After this point the three TNOs fainter than $m_R=23.7$ were dropped
from the followup effort; the CFEPS TNOs would seldom be this 
faint and these objects had been tracked for one month before being dropped due to magnitude considerations only.
The remaining ten TNOs were then
followed through winter 2004, at CFHT, the NOT, the VLT, KPNO 4-m, 
ESO 2.2-m, and Palomar 5-m. 
Each of these recovery observations generally consisted of a triplet 
followed by a nailing image on a following night. 
Except for CFHT in June 2003 and August 2004 (where the images 
were part of the CFHTLS and thus non-targeted), each of the later 
recoveries was the result of a targeted observation. 
Because the dropping of the three faintest objects was not made
on orbital grounds, this small sample is bias free.

\subsection{Optimal CFEPS Observations}

The experience gained with the presurvey was applied to optimize 
CFEPS observations, in cadence, field layout, and exposure time. 
The optimal timing of each astrometric observation is a balance between 
reducing the rate of false candidates and confusion with main-belt asteroids,
uncertainty in the position of the TNO, and determining accurate orbital elements for each TNO as rapidly as possible. This balancing act is most crucial in the first year of observations, from discovery to one-year recovery. 

\subsubsection{Observing cadence}
The discovery triplets are taken at opposition, as the sky motion then allows separation 
between TNOs and asteroids. 
Each image is part of a larger contiguous ``block'' of between 15 and 
20 square degrees, similar to the presurvey field layout and as shown 
in Fig.~\ref{fieldlayout}.  
By grouping pointings into contiguous blocks, we enhance the likelihood of 
recovering the previously-detected TNOs without the need for an accurate 
ephemeris prediction.  
This minimizes the role of the assumptions required to create an
orbit estimate based on short-track TNO observations, which induce 
potential bias in a tracking strategy based on pointed recovery.
Checkup and recovery observations cover the same sky pointings as the 
discovery triplets; by placing the fields contiguously a smaller 
percentage of objects move out of the block of pointings. 
The compact layout of the block also ensures easy cycling through 
fields at the telescope, with a minimum of overhead.

\begin{figure}[htpb]
\includegraphics*[width=7in]{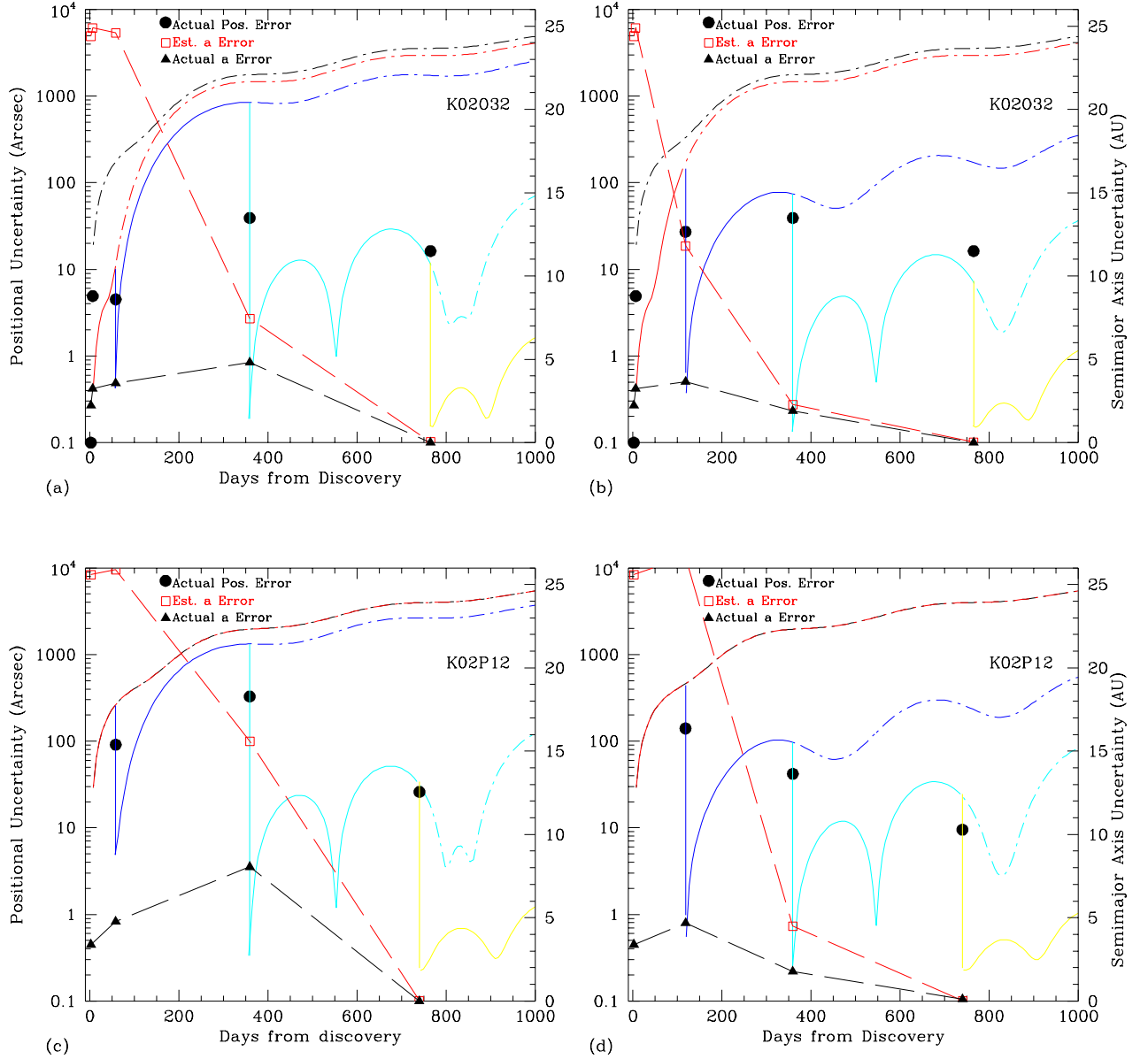}
\caption{ Time evolution of positional and semi-major axis uncertainty for two of our presurvey classical-belt objects. Symbols are as described in Fig.~\ref{CR105}. Panels (a) and (b) show the difference between two month and four month checkup. Panels (c) and (d) are similar, for a different object which did not receive a nailing observation. The uncertainties are generally low for these objects, as they are on nearly circular orbits. \label{K02O32}
}
\end{figure}

The nailing image (of the same field pointing) is taken within 
the next four days so that the positional uncertainty of the object on this single
image is small enough to identify the TNO with a high degree
of certainty. The nailing observation is not strictly necessary in order to find the 
TNO again two months later, as the positional error by that time may 
only grow to $\approx100\arcsec$. However, we find that the nailing observation improves the 1-year ephemeris prediction; with only the discovery triplet and a checkup triplet 
two months later the 1-year ephemeris error can reach one degree,
at which point the object could be considered ``lost''.  The utility of the nailing observation is illustrated 
in Fig.~\ref{K02O32};  with the nailing observation the ephemeris uncertainty at one year is
a factor of two lower.

The checkup observations could in principle be conducted at any time in the next four months before 
the objects disappear behind the Sun. One might think that waiting as long as possible is best, and
indeed Fig.~\ref{K02O32} shows that with a 4-month checkup, 
1-year semi-major axis uncertainties are a factor 3 smaller and
1-year ephemeris uncertainties are reduced by an order of 
magnitude when compared with checkup observations conducted at 2 months. 
Unfortunately, practical considerations intervene.

This is due to the nature of the checkup observations. The CFEPS check-up images are not targeted recoveries but rather
re-imaging the initial survey pointings (due to secondary science goals for
the LS-VW outside the scope of this project, but this also helps avoid orbital bias in tracking). 
Since these pointings 4 months after opposition are available only for
a short time at the start of a night and are moving $<1\arcsec$/hr,
scheduling a survey like CFEPS becomes risky and difficult.
One observation per night at the beginning of several contiguous
nights would be required and this is very difficult to acquire
consistently and to analyze because of non-linear TNO motion over sequential nights.
With CFEPS needing to acquire check-up observations on 15--20 fields
per dark run, waiting until four months after opposition becomes impractical.
Waiting until three months after opposition has similar problems and the TNOs are near the stationary points in their orbits. Therefore, due to practical considerations outside the orbital determination, the CFEPS checkup observations are conducted two months after opposition. As the ephemeris error ellipse is much smaller at two months compared to four, this is also a more conservative observing strategy,

One year recovery observations consist of another triplet, followed by a single 
image on a nearby night, similar to the original discovery sequence. 
These images are taken near opposition, again to avoid asteroid confusion. 
A two-month deviation from this timing, such as would be possible within 
the queue, does not make much of a difference towards reducing uncertainty 
in semi-major axis or positional error, but would make detecting the 
TNOs much more challenging. 
The one-year recovery observation reduces future positional uncertainty to 
the order of an arcminute. However, with a two-month checkup, the semi-major axis uncertainty is still between 1 and 10 AUs after this one-year recovery (see Fig.~\ref{K02O32}). 
A further observation at or after two years is required to reduce the 
orbital parameter uncertainties another order of magnitude, to the 
levels necessary for dynamical modeling or resonance family 
determination. 

This CFEPS observation strategy is designed to minimize TNO losses, and particularly to minimize any orbital bias in which objects are recovered. An insidious problem in TNO surveys has been recovery biases due to orbital assumptions. If the object is searched for on a later date, and not recovered, is it lost because it is below the limiting magnitude, in front of a star or in a chip gap, or because its orbit is actually quite different than predicted, and thus has moved out of the field of view? Because the CFEPS checkup and recovery observations are non-targeted, cover large patches of sky and occur at a rapid time cadence to avoid large ephemeris uncertainties, these objects should not be lost due to assumptions in the orbit. The other factors in losing objects --- chip gaps, bright stars, limiting magnitude problems --- have no correlation with the TNO orbit. Some objects in the survey will be lost out of the field of view, as the CFEPS checkup and recovery fields must observe the same pointings as the discovery fields, due to the Legacy nature of the CFHTLS-VW. The fraction of discovered objects which are not tracked will be characterized for each block.

\subsubsection{Exposure times and filters}

With Megacam, the camera used for CFEPS, the readout is much shorter than with CFH12K; 30 seconds instead of 2 minutes. As such, we are able to use shorter exposure times without dropping below a 50\% duty cycle. While we were able to follow 10 out of 13 presurvey TNOs for 2 years, three of the faintest TNOs were lost after one month. This illustrates the flux bias present in all TNO followup. In particular, it illustrates the difficulty of tracking $R>23.7$ TNOs in variable weather conditions, even with 4-meter and larger telescopes! To avoid this, we change to 70 second exposures for the discovery triplet in CFEPS, with a $g'$ filter. This ensures that almost all TNOs discovered are brighter than $R=23.5$ (assuming $V-R$ = 0.5) and much more easily trackable at a range of telescopes, while still providing a decent duty cycle at CFHT. It also generates a discovery rate of approximately one object per pointing. The nailing image is taken in the same filter, with a slightly lengthened exposure time (90s) to avoid losing TNOs below the limiting magnitude if the seeing varies. 

Grouping field pointings into blocks provides a convenient way to cycle through each triplet. With the lower overhead of Megacam, the typical time spacing between images is only 50 minutes, instead of 1.5 hours. With typical seeing conditions of 0.8\arcsec, this is still adequate to detect the motion of TNOs as distant as 150~AU. 

Checkup is done in $i'$ with a longer exposure time (180s) to detect the faintest TNOs, assuming a $g'$-$i'$ color of approximately 0.8. To provide imaging of the same fields in different filters for other projects, the one year recovery is performed in $r'$, with 110 second exposures in the triplet plus nailing image.  The range of filters is necessitated by secondary science goals of the LS-VW, such as stellar populations and galactic structure studies.

\section{Data Reduction}

Our automated pipeline\citep{Petit04} was used to search each triplet of images for moving objects. The discovery triplets are the images which were searched and characterized using the pipeline, with variants being used to help identify moving objects in checkup and recovery images.

\subsection{The Automated Pipeline}

The presurvey discovery observations from CFH12K were trimmed and bias corrected, then flat-fielded using a median of all the discovery triplets combined with strong rejection criteria to remove stars and galaxies. Each individual flat-fielded image was then implanted with artificial TNOs and searched for moving objects, using two different techniques. Two catalogues of all objects detected in each image were made, one produced using Source Extractor and the other via a wavelet technique. Each of these catalogues for the triplet were then compared, and stationary objects removed. Remaining objects were checked for movements compatible with TNOs in the field, and candidate objects which appeared in both sets of catalogues (Source Extractor and wavelet) were shown to an operator. Further details on the moving object pipeline process can be found in \citet{Petit04}.

In the CFEPS presurvey, thirteen real TNOs were confirmed out of this set of candidates. Each discovery triplet was searched for objects moving between 1 and 6 $\arcsec$/hr at angles between $\pm 23 ^\circ$ with respect to the ecliptic plane. 

The same procedure will be used to find moving objects in each image triplet taken in the LS-VW. The CFEPS TNOs are detected in a much larger field of view (36 CCDs covering 1 square degree instead of 12 CCDs covering 0.32 square degrees), but the techniques have been well tested and scale up to the larger data set nicely. With the exposure times as above, given the TNO luminosity function measured by \citet{Gladman01}, we expect to find approximately one TNO per square degree.

\subsection{Detection Efficiencies}

The detection efficiency of the discovery triplets was calibrated as a function of magnitude and rate. After implanting artificial moving objects and running the automated software, the resulting list of detected objects is compared to the list of implanted objects. It is obvious that the magnitude of a moving object will affect how frequently it is detected, as a lower signal to noise object is more likely to escape the  cataloging process within the moving object software pipeline. 

The rate of motion of an object also affects the detection efficiency. Objects which are moving very rapidly can travel distances of order the seeing disk (or, a more typical threshold of half the seeing disk) during an exposure, leading to ``trailing loss'' as the signal from the object is spread out over a larger number of pixels. Efficiency losses can also occur for objects which are moving very slowly, due to misclassification of moving objects as stationary, or rejection as TNOs due to perceived nonlinearity in their motion. This ``misclassification loss'' affects low signal-to-noise objects more severely than bright objects, due to poorer centroiding in each image. At higher motion rates, the poor centroiding effect becomes less noticeable, as the error drops to a smaller percentage of the measured motion. 

\subsubsection{Presurvey efficiency}

\begin{figure}[htpb]
\includegraphics*[width=7in]{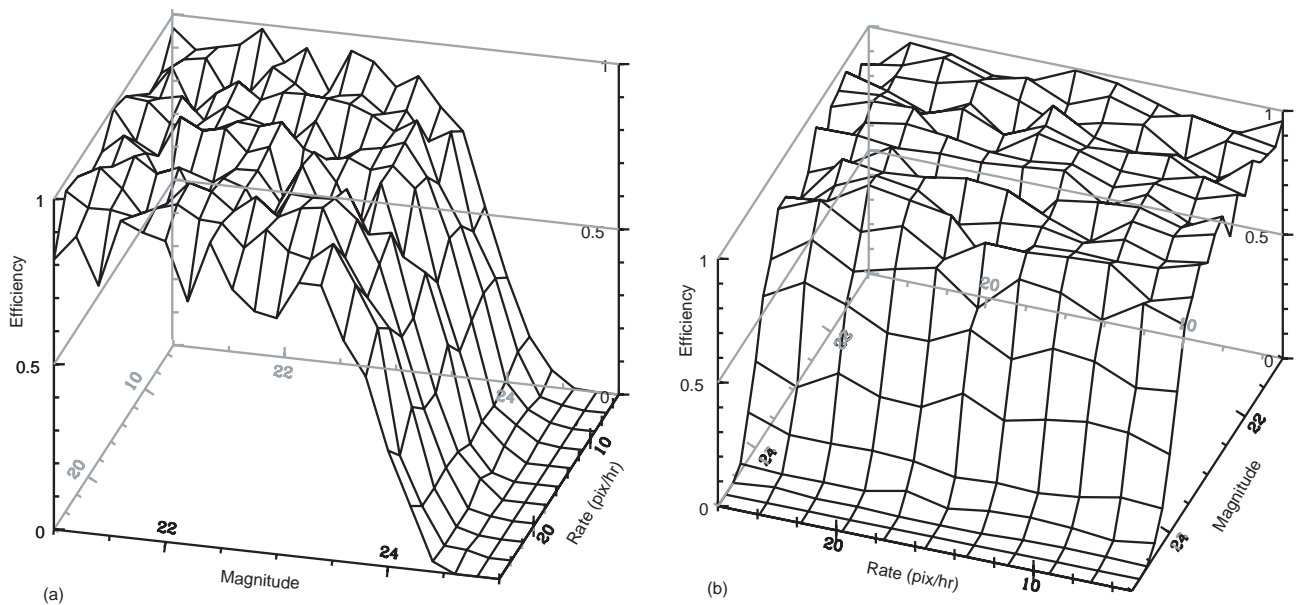}
\caption{Detection efficiencies for the CFEPS Presurvey, split into bins of magnitude and TNO motion rate. Panel (a) is a viewpoint showcasing variation with magnitude --- note the steep drop-off at all rates near $R=24$. This is consistent with previous TNO search efficiency calibrations. The view in panel (b) shows variation with rate (in pixels / hour) for each magnitude bin. Note the slight decrease in efficiency at low rates compared to the efficiency at a higher rate for similar magnitudes.
 \label{efficiencies} }
\end{figure}

We implanted a total of 42286 artificial objects with random motions across all ranges searched in the presurvey, and with a flat magnitude distribution ranging from 21.0 to 25.0. The number of artificial TNOs placed in each chip was approximately 150. Although trailing losses are not an issue for the presurvey, as each exposure was only two minutes in length, we have attempted to measure our detection efficiency in both rate and motion, Fig.~\ref{efficiencies}. Misclassification losses were also negligible, due to the combination of good seeing, long time gap between exposures in the discovery triplet, and searching to a minimum rate of $1\arcsec$/hr.

As shown in Fig.~\ref{efficiencies}, the efficiency is the percentage of objects detected in each magnitude/rate bin compared to the number of objects implanted in the bin. Even though we implanted over 40,000 artificial objects into this relatively small area (6.5 square degrees), the resulting efficiency function is still somewhat coarse. Given that the rate effects are negligible in the presurvey, we have simply fit a smooth double hyperbolic tangent function to the magnitude efficiency falloff: 
\begin{displaymath} 
\eta ({R}) = \frac{A}{4} [1-\tanh{(\frac{R - R_c}{W_1}})]  [1-\tanh({\frac{R - R_c}{W_2}})] \label{efffunc}
\end{displaymath} 
where $A$, $R_c$, $W_1$ and $W_2$ are equivalent to the maximum efficiency, rollover magnitude, and equivalent widths respectively \citep{Petit05}. For the presurvey, $A=0.906$, $R_c=24.20$, $W_1=0.615$, $W_2=0.309$. This places the 50\% efficiency level at $m_R=23.9$ with a very steep falloff beyond this magnitude. Due to the rate cutoffs in the search, the efficiency is zero for rates $<1\arcsec$/hr or $>6\arcsec$/hr.

\subsubsection{CFEPS efficiency}

As the CFEPS is completed in blocks, each block will have an average efficiency function in rate and magnitude. Trailing losses are ignorable for the short exposure times of the CFEPS discovery images, so we will provide a similar smooth double hyperbolic tangent magnitude efficiency function for each block. The detection efficiency tables will also be made available online for CFEPS fields.

\section{The Presurvey Discoveries}

The orbital information for each of the TNOs discovered in this survey is listed in Table~\ref{objectstable}. The orbits and uncertainties are calculated using the \citet{Bernstein00} software, version $2.0$. 

We did not detect any confirmed plutinos in this sample. The presurvey field coverage is only 20 degrees away from Neptune, placing the plutinos near aphelion and at the faintest end of their magnitude range. Thus the expected fraction in the sample is lower and so this lack of plutinos is not entirely surprising. We also did not confirm detections of any Scattered Disk objects (SDOs), although one of the three objects which was lost was potentially an SDO (K02O01, with a high eccentricity and inclination). This leaves our ten well-tracked detections with semi-major axes between 42.6 and 47.5~AU, and distances between 39.1 and 50.6~AU. One of our TNOs, K02O12, appears to be in the 2:1 mean-motion resonance, while two others are potentially in the 7:4 resonance. 

In order to demonstrate our survey simulator, we will consider all of these objects to be ``classical belt'' TNOs. This avoids the problem of adding resonance angles into the orbital model, and with such a small sample, is unlikely to make a significant difference. And so, we now describe in the final section, the survey simulator. 

\section{The CFEPS Survey Simulator}

One of the primary goals of CFEPS is to be able to compare the results
of our observational campaign against theoretical models with high precision.
Therefore much of the design of CFEPS has been focussed on
the need to have sufficient information available so that a
model of the Kuiper Belt can be `passed through' a simulator
of the CFEPS survey. These modeled detections can then be compared
against the objects actually found. 
A statistical comparison of object properties can then be made against any available known parameters,
such as magnitudes, distances, and orbital elements.
In particular, knowledge of the pointing history of
the telescope and the detection efficiency of each block of
the survey is critical. If the magnitude and areal coverage is not
uniform, different populations of Kuiper Belt objects are
favored (for example, plutino detection is much poorer
near conjunction with Neptune than away). An unbiased TNO orbital tracking program, as described above, is also critical. Otherwise dynamical classes could be missed or their relative populations changed (for example, if Extended Scattered Disk Objects were preferentially lost due to incorrect orbit assumptions). 

The content of the CFEPS simulator will evolve as more and more blocks
are added to the CFEPS survey; blocks will be added as their objects
come to orbits with arcs 2 years or longer (the first release will thus 
occur in late 2006). 
In this section we present the general algorithm of the CFEPS
simulator but apply it only to the small pre-survey sample.

Because the orbital parameter space of the Kuiper Belt is vast,
'inverting' the observed objects to an
intrinsic orbital distribution is problematic.
\citet{Bottke00} discuss this in the case of modeling the
near-Earth object (NEO) detections from the Spacewatch telescope's
NEO survey in order to determine the fractional contribution
of several main belt input sources.
The Kuiper Belt problem is somewhat more difficult due to the
fact that we are directly modeling the source region itself
which thus has many more parameters than the simple single
number corresponding to a fractional weighting, as in \citet{Bottke00}. That is, we 
must model the orbital distribution interior to the classical
belt itself, rather than just assign a single weight to the 'classical belt'.

In addition, the importance of resonant interactions in the
observed orbital parameter space requires that the constraints
(coming from the resonant angles) between the various longitudes
of the particles relative to the planets (mostly Neptune) be
respected; TNOs in resonances come to perihelion (where the
strong detection bias favors discovery) at certain ranges
of ecliptic longitudes. In this paper, we will concentrate on modeling the classical
belt, a much simpler problem.

\subsection{Simulator algorithm}

The CFEPS simulator begins with a model distribution of the
Kuiper belt (or Kuiper belt region of interest) and applies
a set of detection constraints related to the known history of the CFEPS
survey (field location, magnitude depth, rate sensitivity)
in order to draw a user-determined number of model detections.
These model detections can then be compared to the real detections
to determine if the model can be rejected statistically.

The Kuiper belt model consists of a file of one of several forms.
The simplest consists a (typically large) number of ($a,e,i$)
triplets, with a specification that their H absolute magnitudes
should be drawn from a power law distribution of the form
\begin{displaymath}
N (H < h) \propto 10^{\alpha h} ,\; H_{min}  <  h < H_{max}
\end{displaymath}
where $H_{min}$ and $H_{max}$ are the minimum and maximum values
and $\alpha$ is the slope of the power law.
Published values for the slope $\alpha$ in the apparent 
magnitude range down to $m_R\simeq 24$ range from 
$ 0.6 \leq \alpha \leq 0.9$ \citep{Trujillo01, Gladman01, Elliot05, Bernstein04}. 

A second possible form for the model distribution allows the
user to specify the H magnitude of each ($a,e,i$) triplet, 
although this will usually necessitate a much larger model
file (to accurately represent the magnitude distribution).

A third possible model distribution is one consisting of a 
six-element set (including the longitudes of node $\Omega$, argument
of pericenter $\omega$, and mean anomaly $M$) where the $H$-magnitudes 
are distributed by the same power law as above; this form
is necessary at present if resonant angles are to be taken into account
as the mean anomaly cannot be chosen independent of $\Omega$
and $\omega$ (note that Neptune's position needs to be taken
into consideration).
Lastly, the $H$ magnitudes may be explicitly added to each
of the 6-element sets instead (allowing different size distributions
for different populations to be taken into account).

The CFEPS pointing history consists of a file with details corresponding
to each block of the survey. 
Each block consists of a characterized region of search, parameterized
by its width and height (in degrees), the RA and declination center 
of the block, the Julian date of the pointing, the filling factor of
the block (what fraction of the area inside the width and height 
specified is not in chip gaps for the Mosaic), the fraction of the discovered sample
in that block which was tracked to $\geq$2-year arcs,
and a pointer to the efficiency function for that block.

The efficiency function for each block contains a definition of the magnitude efficiencies, either given as an analytic fit to the efficiency function (for example, Eqn.~\ref{efffunc} for the presurvey) or as a lookup table. The rate cuts of the detection pipeline can also be specified.
For example, the presurvey discovery data was searched only between
retrograde rates of 1 and 6 $\arcsec$/hour.
The lower rate cut is important in accurately determining the
ability of surveys to detect large distant objects in the 
scattered or extended scattered disks. 

The survey simulator then accepts a request for a certain number
of model detections, and proceeds to iterate until this is
successful or 10 billion iterations have occurred.
The simulator first determines the faintest $H$ magnitude that
could be seen in the faintest of the input blocks.
The looping algorithm is:
\begin{itemize}
\item
Choose a random line from the model distribution. If necessary, determine a
random $H$ magnitude from the specified power law.
\item
If the $H$ magnitude is greater than the faintest possible, redraw.
\item
If not defined, determine random values for $\Omega$, $\omega$,
and $M$, uniform on [$-\pi$,$\pi$]; if these are not uniform
an explicit distribution should be part of the input model.
Calculate the apparent R-band magnitude of the object, given its distance and $H$. 
\item
Determine if the object is both in the field of view and within the rate cuts
of the survey; if not, redraw.
\item
Determine if the object is seen, using the magnitude efficiency function
and a random number from 0--1 to simulate the detection probability; if not seen, redraw.
\item
Based on the `tracking fraction', randomly decide if the object
is tracked, but record non-tracked detections as well.
\item
Record the orbital elements and $H$ magnitude of the detection,
along with apparent magnitude, distance, and sky coordinates.
\end{itemize}

As one can imagine, when the sky coverage is small or the $H$ distribution very steep, many random draws from the model distribution must be made to generate
sufficient model detections.
As the sky coverage and magnitude reach of the surveys grows,
fewer trials are required. Depending on the input Kuiper belt models, the amount of time required to generate a ``simulated survey'' can vary widely. Generating 10,000 detections for the presurvey data requires about 1 hour of CPU time on a 3GHz-class PC for a simple case where objects are always bright enough to be seen in the presurvey field (but the random object does not necessarily fall in the presurvey field of view). A simple power law distribution for H under the same $a$, $e$ and $i$ conditions requires about 200 hours of CPU time.

\subsection{Survey Simulator Examples}

The survey simulator recreates the biases inherent in the survey, allowing users to thoroughly understand how the theoretical model translates to detected TNOs. We will now illustrate the use of the survey simulator and compare the presurvey objects to classical Kuiper belt models, as an example of the future use of the simulator rather than with the expectation of making strong conclusions with our 10-tracked-object sample size.

\subsubsection{Illustrative Examples\label{examples}} 

Let us consider a very simple model with each Kuiper belt object having a fixed semi-major axis $a=44$~AU, and inclination $i=2.0^\circ$ (as the presurvey fields were centered on an ecliptic latitude of 2$^\circ$). In order to illustrate the effects of survey limiting magnitude on detected TNO orbital parameters, we create a series of models where all objects have the same eccentricity, $e=$[0.01, 0.1, 0.2, 0.3], and a power-law absolute magnitude distribution,  $N\propto 10^{\alpha H}$, $\alpha = $[0.6, 0.8, 1.0], $2.0<H<11.0$. For these models, we have chosen to pick 1,000,000 ``detections'' from the simulator in each run --- far above what would actually be discovered in our small field, but useful to increase the signal to noise in the orbital parameters found by the simulated survey. 

\begin{figure}
\includegraphics*[width=5in]{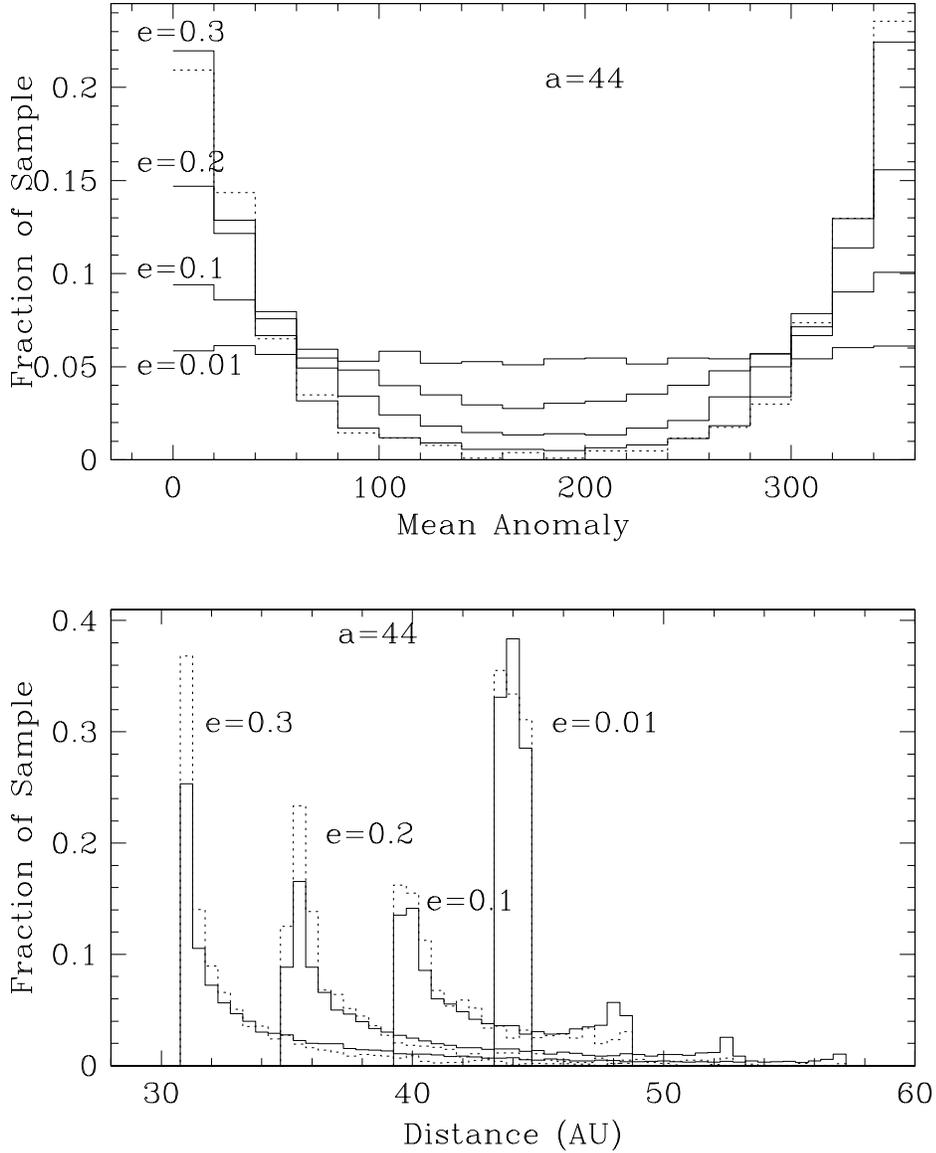}
\caption{The mean anomaly and distance results of toy model simulator runs, with a range of eccentricities and $H$ power-law distributions. Top: the mean anomaly distributions, with $H$ distributions with $\alpha=0.6$ shown as solid lines and labelled with $e$. The dotted line ($\alpha=1.0$, $e=0.2$) illustrates the degeneracy between the $H$ distribution and the eccentricity distribution. Bottom: The distance distribution. $\alpha=0.6$ are shown as solid lines, $\alpha=1.0$ are shown as dotted lines. The mean anomaly and distance distributions above show a bias towards detection at perihelion that strengthens with increasing eccentricity and steepening $H$ slope.
\label{simres1} }
\end{figure}

Looking at the detected distance and mean anomaly distributions produced by the simulations of this toy model (Fig.~\ref{simres1}), we can quantify some interesting effects. As the eccentricity rises, there is a stronger bias towards finding the simulated TNOs at perihelion. This is visible in the mean anomaly as an  ``aphelion hole'' where dramatically fewer objects are detected near mean anomaly $180^\circ$. For a given eccentricity, the depth of the aphelion hole depends on the slope of the $H$ distribution; thus determination of the eccentricity distribution would require a joint fit with the size distribution. The aphelion hole is also visible in the distance distribution as an increasing fraction of objects detected near perihelion. The strength of this perihelion bias is remarkable, and might be surprising if one considers that objects with eccentric orbits spend the majority of their time near aphelion. For our shallowest ($\alpha=0.6$) H-power law distribution of $e=0.3$ TNOs, 49\% of the simulated objects were discovered within 2~AU of perihelion, although they spend only 12.7\% of their orbital period within this distance. With a steeper ($\alpha=1.0$) power law, this fraction increases to 66\% within 2~AU. This is caused by the steep falloff in TNO flux as distance $d$ increases ($\propto d^{4}$) combined with a cutoff in the detection efficiency at some magnitude. Not all individual TNOs will exhibit this bias: for large TNOs which remain above the survey efficiency falloff, there is a larger peak in detection likelihood near aphelion due to the increased fraction of time spent at these distances during the orbit.  However, for any distribution of TNOs where a significant fraction become fainter than the detection efficiency falloff at some point in their orbit, this strong perihelion bias overwhelms the aphelion peak. 

To explore this further, consider a second toy model with $a=44$~AU, but with equal quarters of the model having the four eccentricities listed above, with a power law H distribution $\alpha=0.8$.  Within each eccentricity grouping, there is still the same strong bias towards detecting objects near perihelion. Moverover, this bias towards finding objects at smaller distances is reflected in an overall bias towards detecting high-$e$ objects. Although only 25\% of the input toy model TNOs had $e=0.3$, 44\% of the detected objects were from the $e=0.3$ group (with the rest of the `detections' being 25\% of $e=0.2$, 17\% of  $e=0.1$, and 15\% of $e=0.01$). Due to the power law $H$ distribution, these detection ratios remained constant as the limiting magnitude of the simulated survey was altered, as long as that limiting magnitude did not surpass the limits of the input model magnitude distribution.

\subsubsection{Simulation of Presurvey}

\begin{figure}
\includegraphics*[width=6in]{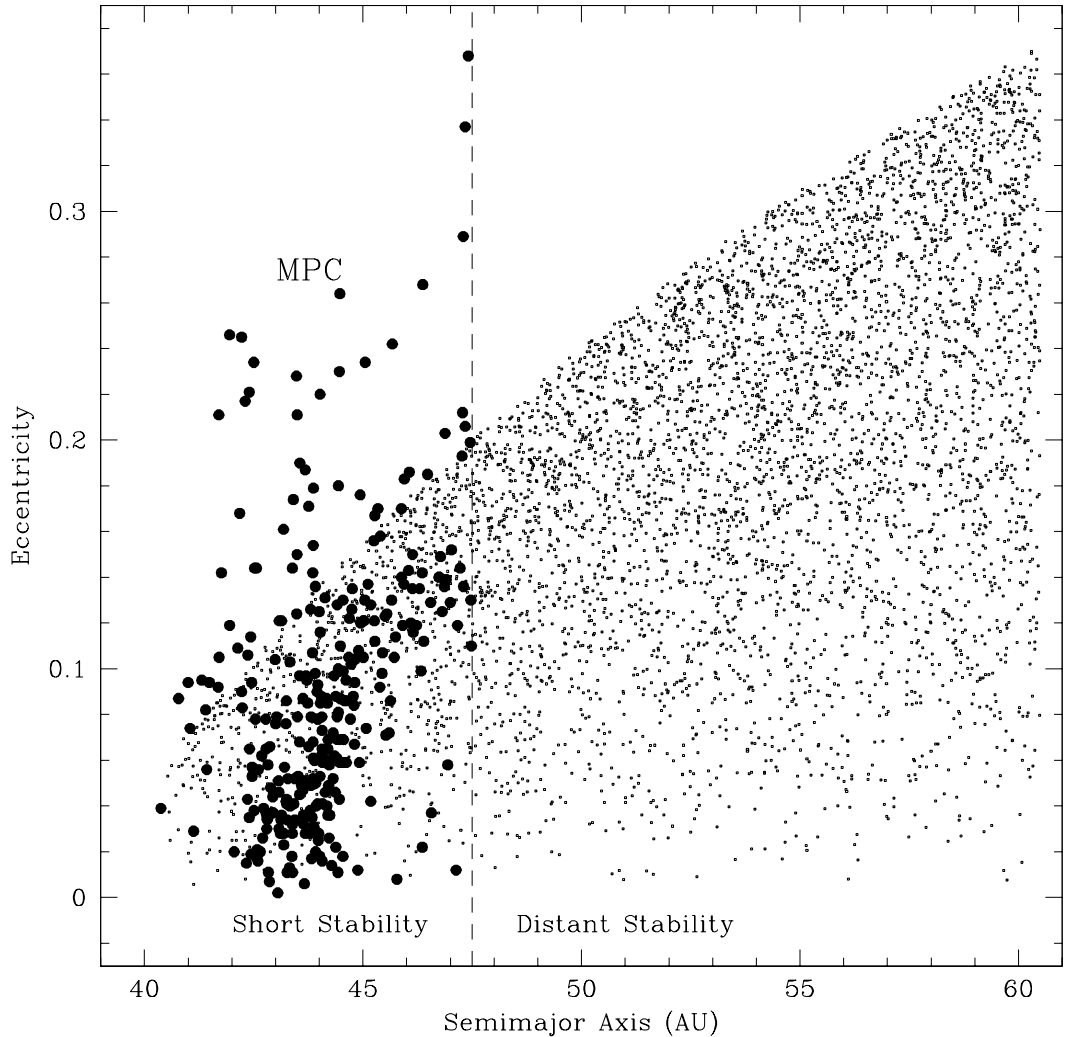}
\caption{Semi-major axes and eccentricities of the models used as input for the simulations of our presurvey discoveries. The solid circles are the objects from the MPC database used as an empirical model, with $40.5<a<47.5$~AU and multi-opposition orbits. The small dots show the stable phase space models used, with $40.5<a<47.5$~AU (short stability model) and $40.5<a<60.5$~AU (distant stability model) subsets used as described in the text.
\label{modelinput}
}
\end{figure}

We now compare our presurvey detections against three more complicated models, representing either empirical or theoretical models of the classical belt region.
For an empirical view of the Kuiper belt, we will consider the Minor Planet Center database of classical TNOs as our model distribution. Due to the generally low eccentricities ($e<0.3$), moderate distances, and easier observational tracking of objects in this region, it is tempting to attempt to use the group of objects in the MPC database as a representative sample of the true population. We therefore take the $a$, $e$ and $i$ of TNOs in the MPC database with $a$ between 40.5 and 47.5~AU and multi-opposition orbits as the orbital input to our simulator (304 objects, plotted in Fig.~\ref{modelinput}). 
As a theoretical view, we note that according to numerical simulations \citep{DLB95}, most of the classical belt with $q>$38~AU is stable for $a>40.5$~AU. We thus build two other simple models by assuming that the classical belt is part of a disk that slowly declines in surface density (falling proportional to $a^-{\frac{3}{2}}$) and was excited by some mechanism which left phase space uniformly filled in $e$; we
populate each semimajor axis up to an $e$ corresponding to $q$=38~AU, but weighted by $e^2$ since there is more phase space volume at
higher $e$ (see Fig.~\ref{modelinput}).   This provides a Kuiper belt model with stable phase space evenly filled. Two versions of this model are considered, one with $40.5<a<47.5$~AU (the ``short stability'' model) and one with $40.5<a<60.5$~AU (the ``distant stability'' model), to test the presence of a cutoff near the 2:1 resonance at 48.4~AU. 

While we would hope to be able to determine between the obvious difference of a model truncated at 47.5~AU and a similar model extended to 60~AU, we would also like to distinguish between the short stability model and the MPC database of classical TNOs. 
The existence of an `edge' to the Kuiper belt near 47.5~AU has been recognized in previous work \citep{Dones97, JLT98, Allen02, Trujillo01, TrujilloBrown01, Petit05}, but this has previously been based on statistical analysis of the distance distribution only. The existence and location of such an edge has important implications for models of the formation of the Kuiper belt (e.g. \citet{Hahn99, Gomes04}). The difference between the MPC database and the theoretical model is more subtle (see Fig.~\ref{modelinput}); the MPC database shows a lack of low-$e$ classical TNOs with $q>44$~AU when compared to the stability model. It is possible that the lack of low-$e$ detections in the MPC database is due to the decreased detectability of these high $q$ objects, an effect which would be re-created in the survey simulator. The relative number of TNOs in this region can be used as a discriminator between models of the evolution of the outer Solar System (e.g \citet{Gomes03, Levison03}).

We randomize the longitude of perihelion, the longitude of node, and the mean anomaly and pick 50000 objects from the simulator for each model within each group, setting the H power law distribution to be $\alpha=$ [0.6, 0.8, 1.0]. Thus, we ran 9 simulations. 

\begin{figure}
\includegraphics*[width=7in]{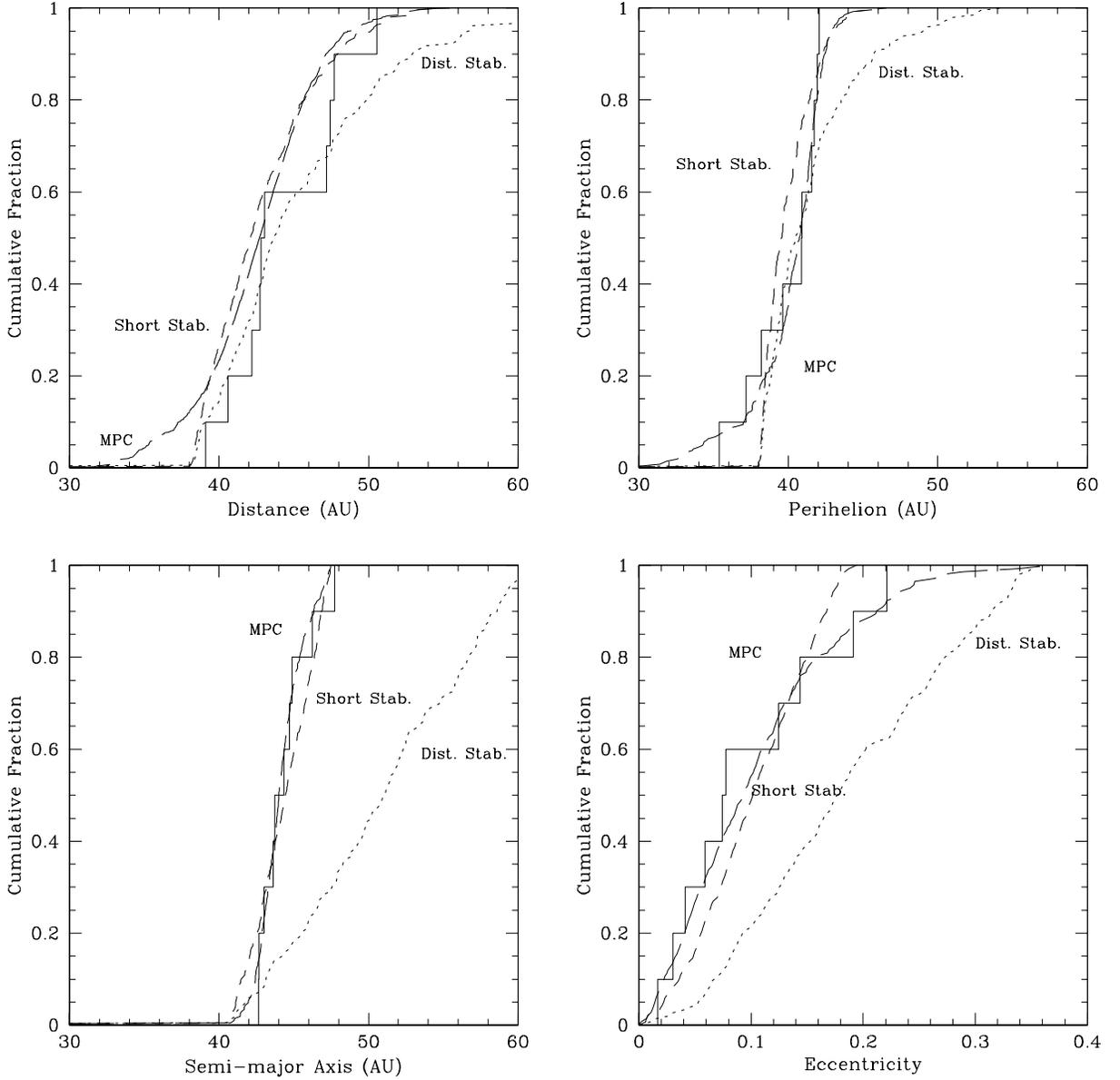}
\caption{The cumulative distance, perihelion, semi-major axis, and eccentricity distributions for the presurvey objects (the stepped solid line) and the simulation results. The simulation objects are the results of running an MPC Classical Disk-based model with $\alpha=0.8$ through the simulator (long-dashed line labeled  ``MPC'' ),  a stable phase space model with $\alpha=0.8$ and $40.5<a<47.5$~AU (dashed ``Short Stability'' line), and an extended stable phase space with $\alpha=0.8$ and $40.5<a<60.5$~AU (dotted ``Distant Stability'' line).
\label{simres2}}
\end{figure}

The cumulative distance distributions from the simulations can be compared against the 10 presurvey detections tracked for 2 years. We conducted a K-S test by calculating the maximum difference between the cumulative distance distribution of each simulation and the cumulative distance distribution of our survey. For each model, the likelihood of this difference was then calibrated by drawing 1000 random samples of 10 objects from the 50000 simulated survey discoveries, and calculating the differences between their cumulative distance distributions and the total simulation distribution. The results of our calibrated K-S test are shown in Table~\ref{modeltable}, and the $\alpha=0.8$ cumulative distributions of these models are shown in Fig.~\ref{simres2}. Table~\ref{modeltable} shows that our small sample does not rule out (P$<5\%$) any of these models based on the distance distribution alone, except for the $40.5<a<47.5$~AU stability model with $\alpha=1.0$.

The power of the survey simulator, when coupled with a quantitatively characterized survey, is in the ability to compare a wide range of model parameters against the true survey results. As such, we can look not just at the distance distribution, but also at eccentricity distributions, semi-major axis distributions, or the perihelion distribution. Table~\ref{modeltable} contains K-S test results in these orbital parameters. The semi-major axis and eccentricity distributions are poor at constraining the two sets of models which are limited to semi-major axis between $40.5<a<47.5$~AU, due to our limited sample size. However, the semi-major axis distribution expected from the distant stability model should have half of the detections with $a>50$~AU, which is very different than the  presurvey distribution where the largest semi-major axis was $a=47.7$~AU. The eccentricity distribution is also quite different, as the eccentricities of our presurvey objects are lower than the simulation results for a distant disk. This is due to the fact that we are mainly ``discovering'' TNOs with $a>50$~AU and high-$e$ when they are near perihelion in the simulation. It also is why the same model ($40.5<a<60.5$~AU) cannot be rejected based on the perihelion or distance distributions of our small sample, due to the strong perihelion bias explained in section~\ref{examples}.
The perihelion distribution can be used to rule out the $40.5<a<47.5$~AU stability model with $\alpha=1.0$ but this is influenced by our assumed $q=38$~AU cutoff which does not account for TNOs in resonances which can have smaller perihelia nor the existence of scattered disk objects.

We find that we can rule out a stable phase space model with $40.5<a<60.5$~AU through the semi-major axis and eccentricity distributions of our presurvey objects. With a larger sample size, more stringent limits could be placed on the presence of a ``cold'' low-$e$ disk outside the 2:1 resonance; as the perihelion distribution in Fig.~\ref{simres2} shows, almost all the $a>50$~AU detections have perihelia inside 50~AU -- the low-$e$ TNOs beyond the 2:1 resonance will make up a very small fraction of the detected objects in a flux-limited sample. 

The distance and perihelion distributions rule out a stable phase space model truncated at $47.5$~AU with a steep TNO $H$ distribution of $\alpha=1.0$. However, we are unable to rule out a $40.5<a<47.5$~AU stability model or an empirical MPC-based model with a shallower magnitude distribution. This is interesting, indicating the lack of low-$e$ classical TNOs with $q>44$~AU in the MPC may just be an observational artifact and the classical Kuiper belt could indeed uniformly fill all stable phase space. 

As a very simple test of what would be possible with a larger sample size, we simply cloned each of our presurvey objects and repeated the K-S probability testing. If a sample twice as large had the same $d$, $q$, $a$ and $e$ distributions, it would be possible rule out all $H$-magnitude distributions of the MPC model based on the distance distribution, all the $40.5<a<60.5$~AU stability models based on the $a$ and $e$ distributions, and all except the $\alpha=0.6$, $40.5<a<47.5$~AU stability model based on the perihelion distributions. 

\section{The future of the CFEPS}

The CFEPS is being carried out using CFHT's Megacam, covering 1 square degree per pointing. One obvious benefit is the huge areal coverage possible in a short amount of time ($\approx 20$ sq deg in 3 hours to $m_g=23.5$). This increases the discovery rate above previous survey levels, but it also increases the percentage of TNOs found in checkup and recovery, as the objects which shear out of the field of view become a smaller fraction of the total discoveries. In the first year of CFEPS results, we have found 65\% of TNOs discovered within the survey can be followed to one-year arcs without using other telescopes. Using other facilities has allowed some blocks to obtain 100\% recovery levels.

We have settled on an observing strategy of discovery triplet, nailing, checkup at two months, followed by recovery at one year. The final measurement of semi-major axis and eccentricity will be obtained from a final recovery at either two or three years from initial discovery (depending on the whether the final observations for a given TNO arc are obtained at CFHT or another telescope). This appears to be an optimal balance of practical considerations and scientific gain. 

At present, over 100 TNOs have been discovered and tracked to one-year recovery within CFEPS, but these have still not been followed long enough to measure the orbital parameters accurately.  With the presurvey results presented here, we are able to rule out a simple model of a uniformly filled stable phase space from $40.5<a<47.5$~AU ($q>38$~AU) if the magnitude distribution follows $\alpha=1.0$ and any uniformly filled stable phase space models extending $40.5<a<60.5$~AU with any $H$-magnitude distribution. Using the survey simulator, we are able to compare orbital distribution models with survey results, allowing detailed testing of models. By 2007, the CFEPS will be capable of detailed model testing, and we expect to be able to rule out the majority of models in the literature, allowing the community to refine the remaining (and newly proposed) models of the formation of the outer Solar System.

\begin{figure}[htpb]
\includegraphics*[width=6.3in]{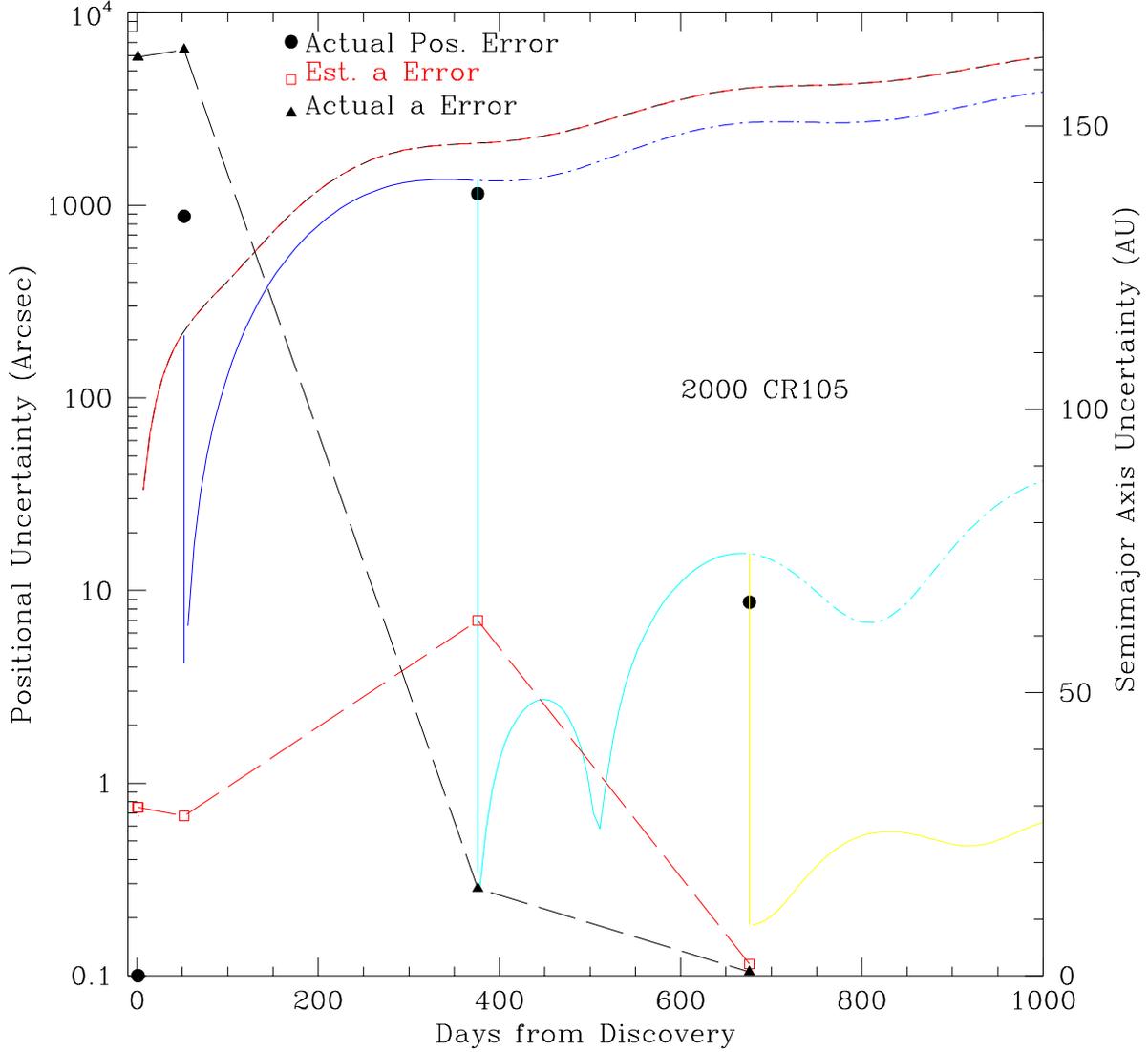}
\caption{Time evolution of positional and semi-major axis uncertainty, for extended scattered disk object 2000 CR105. On a log-normal scale, solid lines indicate the growth of positional uncertainty with time, extended with a dotted line after further observations (at the vertical lines). The measured errors between the expected position and the actual position are indicated by the solid circles. On a normal-normal scale, the straight broken lines illustrate the evolution of the semi-major axis uncertainty. Open squares show the $a$ estimated uncertainty, as given by the Bernstein \& Kushalani code. Solid triangles show the difference between the calculated semi-major axis at each observation and the final semi-major axis measurement. Note the much larger $a$ errors compared with the previous examples (Fig.~\ref{K02O32}).
\label{CR105}
}
\end{figure}


\clearpage

\begin{center}
\begin{deluxetable}{lcccccccc}
\tablecolumns{9}
\tablewidth{0pt}
\tablecaption{Objects Discovered in the CFEPS Presurvey.}
\tablehead{
\colhead{Name} & \colhead{Mag ($R$)} & \colhead{D\tablenotemark{a} (AU) } & \colhead{a(AU)}
& \colhead{e}  & \colhead{i ($^\circ$)} & \colhead{Node($^\circ$)\tablenotemark{b}}  & \colhead{Arg of Peri.} & \colhead{Time of Peri.} \\ 
}
\startdata     
K02O20 & 22.3 & 42.2 & 42.648 & 0.016 & 1.271 & 88.37 & 291 &  2466784 \\
&      & & $\pm$ 0.009 & $\pm$  0.001 & $\pm$ 0.001 & $\pm$ 0.04 & $\pm$ 5 & $\pm$  1272 \\
K02O32 & 22.5  & 47.4 & 44.88 &  0.074 &  3.933 & 339.814 & 206 &  2410815 \\
&      & & $\pm$ 0.02 & $\pm$ 0.002 & $\pm$  0.001 & $\pm$0.004 & $\pm$ 2 & $\pm$ 734 \\
K02O42 & 22.6 & 40.6 & 43.63 & 0.124 & 5.450 & 336.79 & 289.24 & 2437819 \\
&      & & $\pm$ 0.02 & $\pm$ 0.001 & $\pm$ 0.001 & $\pm$ 0.001 & $\pm$ 0.30 & $\pm$ 47 \\    
K02P32 & 23.1 & 42.8 & 42.65 & 0.041 & 1.570 & 119.105 & 112 & 2426450 \\
&      & & $\pm$ 0.03 & $\pm$ 0.004 & $\pm$ 0.001 & $\pm$ 0.023 & $\pm$ 1 & $\pm$ 354 \\
K02P41 & 23.2 & 47.7 & 44.34 & 0.077 & 2.510 & 109.59 & 28 & 2502391 \\
&      & & $\pm$ 0.01 & $\pm$ 0.002 & $\pm$ 0.001 & $\pm$ 0.020 & $\pm$ 6 & $\pm$ 1818 \\
K02P12 & 23.2 & 50.6 & 46.25 & 0.144 & 3.703 & 328.776 & 134.4 & 2492063 \\
&      & & $\pm$ 0.01 & $\pm$ 0.001 & $\pm$ 0.001 & $\pm$ 0.000 & $\pm$ 0.5 & $\pm$ 161 \\
K02O03 & 23.5 & 39.09 & 43.72 & 0.191 & 0.751 & 182.71 & 77.6 & 2438640 \\
&      & & $\pm$ 0.02 & $\pm$ 0.001 & $\pm$ 0.001 & $\pm$ 0.04 & $\pm$ 0.1 & $\pm$ 12 \\
K02O12 & 23.5 & 47.2 & 47.75 & 0.221 & 1.918 & 329.086 & 97.48 & 2477337 \\
&      & & $\pm$ 0.02 & $\pm$ 0.001 & $\pm$ 0.001 & $\pm$  0.001 & $\pm$ 0.06 & $\pm$ 33 \\
K02O43 & 23.5 & 42.7 & 44.702 & 0.059 &  3.574 &  336.730 & 309 & 2440637 \\
&      & & $\pm$ 0.007 & $\pm$ 0.001 & $\pm$ 0.001 & $\pm$ 0.002 & $\pm$ 1 & $\pm$ 278 \\
K02O40 & 23.7 & 43.0 & 43.015 & 0.030 & 3.017 & 110.81 & 125.0 &  2426965 \\
&      & & $\pm$ 0.009 & $\pm$ 0.003 & $\pm$ 0.001 & $\pm$ 0.01 & $\pm$ 0.3 & $\pm$ 95 \\
K02O41 & 23.8 & 42 & 48 & 0.2 & 1.27 & 69 & 190 & 2436197 \\
&   & 2\tablenotemark{c} & $\pm$ 25 & $\pm$ 0.6 & $\pm$ 0.01 & $\pm$ 3 & $\pm$ 100 & $\pm$ 20529 \\
K02O10 & 23.8 & 42 &46 & 0.5 & 3 & 120 & 317 & 2469290 \\
&      & 7 & $\pm$ 20 & $\pm$ 1 & $\pm$ 2 &$\pm$ 23 & $\pm$ 102 & $\pm$ 12142 \\
K02O01 & 23.9 & 59 & 40 & 0.2 & 1.0 & 10 & 130 & 2494782 \\
&      & $\pm$ 5 & $\pm$ 15 & $\pm$ 0.5 & $\pm$ 0.2 & $\pm$ 12 & $\pm$ 100 & $\pm$ 22379 \\
\enddata
\label{objectstable}
\tablenotetext{a}{Distance from Earth.}
\tablenotetext{b}{These orbital parameters are as reported by \citet{Bernstein00}.} 
\tablenotetext{c}{Uncertainty in D is only listed for objects with observed arc $< 30$ days. Other $\Delta$D are $<0.003$~AU. }
\end{deluxetable}
\end{center}

\begin{center}
\begin{deluxetable}{lcccccccccccccccc}
\rotate
\tablecolumns{15}
\tablewidth{0pt}
\tablecaption{Object Tracking}
\tablehead{
\colhead{Object}  & \colhead{CFHT\tablenotemark{a}} & \colhead{C.A.}  & \colhead{NOT}  & \colhead{VLT}    & \colhead{CFHT}  & \colhead{C.A.} & \colhead{Pal}    & \colhead{CFHT}   & \colhead{CFHT}   
     & \colhead{ESO}          & \colhead{VLT}      &  \colhead{CFHT}         & \colhead{KP}    & \colhead{Pal} \\
  &  \colhead{8/02} & \colhead{9/02} & \colhead{9/02} & \colhead{9/02} & \colhead{9/02} & \colhead{10/02} & \colhead{11/02} & \colhead{11/02} & \colhead{6/03}  & \colhead{7/03} & \colhead{10/03} & \colhead{8/04} & \colhead{9/04} & \colhead{9/04} \\
}
\startdata
K02O20   &  5   & 3,4,8 & 3,4 &  --   &   30 & 3   &  --  &   --     &  23            &   28,29                 &   --           &     12        & --   &  -- \\
K02O32   &  5,9 & 8,9    & 4  &  --    &   30 & 3,4  &  --  &   28,29,30 &  23    &   28,29                 &   --           &     --            &    6,7 & --\\
K02O42    &  5   & 3,6,7 &  3,4 &  --     &   30 & 3    & --   &        --    &   24,25         &   28,29              &   --           &      --       &    6,7  & -- \\
\hline 
K02P32  &  5,9 &   --  &  4   & --     &   30  &   --     &  --  &     --       &   24,25           &   29                  &   --           &  --               &    7 & -- \\
K02P41   &  5,9 & 6,7 & --        &  --    &   30  &      -- &  --  & --           &   --                  &   29                   &   --           &       --          &    11  &  16 \\
K02P12  &  5   &     -- &  4   & --     &   30  &    --    & --  &   28,29,30 &   23                  &   28,29                   &   --           &     12        &   --   & -- \\
\hline
K02O03   &  5,8 &  -- &  4   &  -- &  offW   &  -- &  5,6  & -- &    --                  &   --                   &   28,29        &     12        &   -- & --  \\
K02O12   &  5,8 & --    &  4   &  --    &   30 &  --       &   -- &    --       &       --                &   --                   &   29           &     12        &   -- & --  \\
K02O43   &  5,9 &  --   &    --  &  4 &   30 & --       & --   &     --       &    --                      &  --                 &  28,29       &    --             &    11,12 &  -- \\
\hline
K02O40   &  5,9 &  --     &    --    &  4 &  offS  &  --     & --   &     --      &     --                  &   --                   &   29           &      --           &   -- &   16,17 \\
K02O41 &  5,9 &    --  &     --    &  4 &  gap   & --     &  --  &     --      &       --                &   --                   &   --           &         --        &     --   & --\\
K02O10 &  5,8 &    --  &    --     &  4 &  offW   & -- &  --  &        --     &        --                &   --                   &   --           &         --        &     --   & --\\
K02O01   &  5,8 &    --  & --        &  4 &  offW   & --     & --   &     --      &      --                   &   --                   &   --           &     --            &     --   & -- 
\enddata 
\tablenotetext{a}{Telescope abbreviations as follows: CFHT = CFHT 3.6-m + 12K camera, C.A. = Calar Alto 2.2-m, NOT = Nordic Optical Telescope 2.5-m + MOSCA, VLT = VLT + FORS1, CFHT (after 2003) = CFHT + Megacam, Pal = Palomar 5-m, ESO = ESO 2.2-m + WFI, KP = Kitt Peak 2.2-m and 4-m + Mosaic. } 
\label{objecttracking}
\end{deluxetable}
\end{center}

\begin{center}
\begin{deluxetable}{c|ccccc}
\tablecolumns{4}
\tablewidth{0pt}
\tablecaption{Calibrated KS Test Results from comparison of simulated survey ``detections'' and presurvey detections. Sample size is 10 objects.}
\tablehead {  Parameter  & \multicolumn{3}{c}{H power law Index}  \\
& \colhead{$\alpha=0.6$} & \colhead{$\alpha=0.8$} & \colhead{$\alpha=1.0$} 
}
\startdata
& \multicolumn{5}{l}{MPC model, $40.5<a<47.5$~AU}\\    
Distance 			& 18.2\%  & 14.7 \%  & 11.3\%   \\
Perihelion			& 72.1\%  & 81.1\%  & 66.4\%   \\
Semi-major Axis     & 79.5\% & 71.6\% & 58.4\% \\
Eccentricity	     & 63.6\% & 66.9\% & 75.3\% \\
\hline
& \multicolumn{4}{l}{Stability model, $40.5<a<47.5$~AU} \\
Distance 		& 45.6\%  & 15.2\%  & 2.4\%  \\
Perihelion		& 14.2\%  & 10.8\%  & 1.5\% \\
Semi-Major Axis		    & 51.8\% & 64.1\% & 62.4\% \\
Eccentricity			     & 55.5\% & 42.9\% & 13.8\% \\
\hline
&\multicolumn{4}{l}{Stability model, $40.5<a<60.5$~AU} \\
Distance			& 61.6\%  & 87.0\%  & 40.2\%  \\
Perihelion			& 18.5\%  & 22.5\%  & 36.2\% \\
Semi-major Axis	     & 0.0\% & 1.4\% & 0.0\% \\
Eccentricity		     & 1.5\% & 1.4\% & 3.0\% \\
\enddata
\label{modeltable}
\end{deluxetable}
\end{center}

\end{document}